\documentclass[aps,pre,reprint,showpacs]{revtex4-1}

\usepackage{graphicx}
\usepackage{bbm}
\usepackage{url}
\usepackage{bm}
\usepackage{epstopdf} 
\usepackage{amsfonts}
\usepackage{color}
\usepackage{amssymb}
\usepackage{amsmath}
\usepackage{amsthm}
\usepackage{multirow}
\usepackage[unicode=true,pdfusetitle, bookmarks=true,bookmarksnumbered=false,bookmarksopen=false, breaklinks=false,pdfborder={0 0 1},colorlinks=true, linkcolor = blue, citecolor = blue]{hyperref}

\newcommand{\ud}{\mathrm{d}}

\begin{document}

\title{Information geometric duality of $\phi$-deformed exponential families}
\author{Jan Korbel$^{1,2}$}
\author{Rudolf Hanel$^{1,2}$}
\author{Stefan Thurner$^{1,2,3,4,}$}
\email{stefan.thurner@meduniwien.ac.at}
\affiliation{$^1$Section for Science of Complex Systems, Medical University of Vienna, Spitalgasse 23, 1090 Vienna, Austria}
\affiliation{$^2$Complexity Science Hub Vienna, Josefst\"adterstrasse 39, 1080 Vienna, Austria}
\affiliation{$^3$Santa Fe Institute, 1399 Hyde Park Road, Santa Fe, NM 87501, USA}
\affiliation{$^4$IIASA, Schlossplatz 1, 2361 Laxenburg, Austria}


\date{\today}

\begin{abstract}
In the world of generalized entropies---which, for example, play a role in physical systems with
sub- and super-exponential phasespace growth per degree of freedom---there are two ways for
implementing constraints in the maximum entropy principle: linear- and escort constraints.
Both appear naturally in different contexts. Linear constraints appear e.g. in physical systems,
when additional information about the system is available through higher moments.
Escort distributions appear naturally in the context of multifractals and information geometry.
It  was shown recently that there exists a fundamental duality that relates both approaches
on the basis of the corresponding deformed logarithms ({\em deformed-log duality}).
Here we show that there exists another duality that arises in the context of information geometry, relating the
Fisher information 
of $\phi$-deformed exponential families that correspond to linear constraints (as studied by J. Naudts),
with those that are based on escort constraints (as studied by S.-I. Amari).
We explicitly demonstrate this {\em information geometric duality} for the case of $(c,d)$-entropy
that covers all situations that are compatible with the first three Shannon-Khinchin axioms,
and that include Shannon, Tsallis, Anteneodo-Plastino entropy, and many more
 as special cases.
Finally, we discuss the relation between the deformed-log duality and the information geometric duality,
 and mention that the escort distributions arising in the two dualities are generally different and only coincide
for the case of the Tsallis deformation. 
\end{abstract}


\maketitle

\section{Introduction}

Entropy is one word for several distinct concepts \cite{tmh17}. It was originally introduced in thermodynamics, then in statistical physics, information theory, and last in the context of statistical inference. One important application of entropy in statistical physics, and in statistical inference in general, is the {\em maximum entropy principle}, which allows us to estimate probability distribution functions from  limited information sources, i.e. from data \cite{jaynes57,harremoes01}. The formal concept of entropy was generalized to also account for power laws that occur frequently in complex systems \cite{tsallis88}. Literally dozens of generalized entropies were proposed in various contexts, such as relativity \cite{kaniadakis02}, multifractals \cite{jizba04}, or black holes \cite{tsallis13}; see \cite{thk-book} for an overview. All generalized entropies, whenever they fulfil the first three Shannon-Khinchin axioms (and violate the composition axiom) are special cases of the $(c,d)$-entropy asymptotically \cite{ht11a}. Generalized entropies play a role for non-multinomial, sub-additive systems (whose phasespace volume grows sub-exponentially with the degrees of freedom) \cite{tsallis05,ht11b}, and for systems, whose phasespace grows super-exponentially \cite{jensen18}. All generalized entropies, for sub-, and super-exponential systems, can be treated within a single, unifying framework \cite{korbel18}.

 With the advent of generalized entropies, depending on context, two types of constraint are used in the maximum entropy principle:  traditional linear constraints (typically moments), \mbox{$\langle E \rangle = \sum_i p_i E_i$}, motivated by physical measurements, and the so-called {\em escort constraints}, \mbox{$\langle E \rangle_u = \sum_i u(p_i) E_i/\sum_i u(p_i)$}, where $u$ is some nonlinear function. Originally, the later were introduced with multifractals in mind \cite{tsallis88}. Different types of constraint arise from different applications of relative entropy. While for physics-related contexts (such as thermodynamics) linear constraints are normally used, in other applications, such as non-linear dynamical systems or information geometry it might be more natural to consider escort constraints. The question about their correct use and the appropriate form of constraints has caused a heated debate in the past decade \cite{tsallis98,abe03,htt09,htt09a,ohara10,bercher12}. To introduce escort distributions in the maximum entropy principle in a consistent way, two approaches have been discussed. The first \cite{tsallis03} appears in the context of deformed entropies that are motivated by superstatistics \cite{beck03}. It was later observed in \cite{htg12} that this approach is linked to other deformed entropies with linear constraints through a fundamental duality (deformed-log duality), such that both entropies lead to the same functional form of MaxEnt distributions. The second way to obtain escort distributions was studied by Amari et al., and is motivated by information geometry and the theory of statistical estimation \cite{amari10,amari12}. There, escort distributions represent natural coordinates on a statistical manifold \cite{amari12,ay17}.

In this paper, we show that there exists an another duality relation between this information geometric approach with escort distributions, and an approach that uses linear constraints. The relation can be given a precise information geometric meaning on the basis of the Fisher information. We show this in the framework of $\phi$-deformations \cite{naudts02,naudts04,naudts11}. We establish the duality relation for both cases in the relevant information geometric quantities.  As an example, we explicitly show the duality relation for the class of $(c,d)$-exponentials, introduced in \cite{ht11a,ht11b}. Finally, we discuss the relation between the {\em deformed-log duality} and the \emph{information geometric duality}, and show that these have fundamental differences. Each type of duality is suitable for different applications.

Let us start with reviewing central concepts of (non-deformed) information geometry, in particular relative entropy and its relation to the exponential family through the maximum entropy principle. Relative entropy, or Kullback-Leibler divergence, is defined as
\begin{equation}
	D_{KL}(\bm{p} \| \bm{q}) = \sum_i p_i \log\left(p_i/q_i\right) \quad .
\end{equation}
For the uniform distribution $\bm{q}=\bm{u}_n$ i.e, $q_i = 1/n$, we have
\begin{equation}
	D_{KL}(\bm{p} \| \bm{u}_n) = \log n - S(\bm{p}) \quad ,
\end{equation}
where $S(\bm{p})$ is Shannon entropy, $S(\bm{p}) = - \sum_i p_i \log p_i$. It is maximized by the exponential family of distributions. Consider a normalization constraint, $\sum_i p_i= 1$, and a set of constraints, $\sum_j p_j E_{ij} = \langle \bm{E}_i \rangle$. Further, consider a  parametric family with parameter vector $\bm{\theta} \in \mathcal{M}$, where $\mathcal{M}$ is a parametric space. For simplicity, we use discrete probabilities. The family of probability distributions for configuration vector, $\bm{E}_i$, that maximizes Shannon entropy, can be written as
\begin{eqnarray}\label{eq:exp}
	p_i(\bm{\theta}) \equiv p(\bm{E}_i;\bm{\theta}) = \exp\left(\Psi(\bm{\theta}) +  \bm{\theta} \cdot  \bm{E}_{i} \right)  \nonumber \\ = \exp\left(\Psi(\bm{\theta}) +  \sum_j \theta_j  E_{ij} \right) \quad .
\end{eqnarray}
$\Psi(\bm{\theta})$ guarantees normalization. This family of distributions is called the \emph{exponential family}.
Fisher information defines the metric on the parametric manifold $\mathcal{M}$ by taking two infinitesimally separated points, $\bm{\theta}_0$, and $\bm{\theta} = \bm{\theta}_0 + \delta \bm{\theta}$, and by expanding $D_{KL}(\bm{p}(\bm{\theta}_0) \| \bm{p}(\bm{\theta}))$,
\begin{equation}
	g^F_{ij}(\bm{\theta})
	=  \left .  \frac{\partial^2 D_{KL}(\bm{p}(\bm{\theta}_0) \| \bm{p}(\bm{\theta}))}{\partial \theta_i \partial \theta_j} \right |_{\bm{\theta}
	=\bm{\theta}_0} \quad .
\end{equation}
Let us assume a probability simplex, $S^n$, with $n$ independent probabilities, $p_i$, and probability $p_0$. Its value is not independent, but determined by the normalization condition, $p_0 = 1- \sum_i p_i$. For the exponential family it is a well-known fact that Fisher information is equal to the inverse of the probability in Eq. \eqref{eq:exp}
\begin{equation}
	g^F_{ij}(\bm{\theta}) \equiv g^F_{ij}(\bm{p}(\bm{\theta})) = \frac{1}{p_i}\delta_{ij}+\frac{1}{p_0} \quad .
\end{equation}

\section{Deformed exponential family}
We briefly recall the definition of $\phi$-deformed logarithms and exponentials as introduced by Naudts \cite{naudts02}. The deformed logarithm is defined as
\begin{equation}
	\log_\phi(x) = \int_{1}^{x} \ud y \frac{1}{\phi(y)} \quad ,
\end{equation}
for some positive, strictly increasing function, $\phi(x)$, defined on $(0,+\infty)$.  Then, $\log_\phi$ is an increasing, concave function with $\log_\phi(1) = 0$. $\log_\phi(x)$ is negative on $(0,1)$, and positive on $(1,+\infty)$.  Naturally, the derivative of $\log_\phi(x)$ is $1/\phi(x)$.  The inverse function of $\log_\phi(x)$ exists; we denote it by $\exp_\phi(x)$. Finally, the $\phi$-exponential family of probability distributions is defined as a generalization of Eq. \eqref{eq:exp}
\begin{equation}
	p_i(\bm{\theta}) = \exp_\phi \left(\Psi(\bm{\theta}) +  \sum_j \theta_j  E_{ij} \right) \quad .
\end{equation}
We can express $\Psi(\bm{\theta})$ in the form
\begin{equation}\label{eq:psi}
	 \Psi(\bm{\theta}) = \log_\phi(p_i(\bm{\theta})) -  \bm{\theta} \cdot \bm{E}_{i} \quad ,
\end{equation}
which allows us to introduce dual coordinates to $\bm{\theta}$. This is nothing but the Legendre transform of $\Psi(\bm{\theta})$, which is defined as
\begin{equation}
\varphi(\bm{\eta}) = \bm{\eta} \cdot \bm{\theta} -  \Psi(\bm{\theta}) \quad ,
\end{equation}
where
\begin{equation}
\bm{\eta} = \nabla{\Psi(\theta)}\quad .
\end{equation}
Because
\begin{equation}
	\partial_{\theta_j} p_i(\bm{\theta}) = \exp_\phi'\left(\Psi(\bm{\theta}) +  \bm{\theta} \cdot  \bm{E}_{i}\right) \left( \partial_{\theta_j} \Psi(\bm{\theta}) +   E_{ij}\right)
\end{equation}
holds, and using $\sum_i \partial_{\theta_j} p_i(\bm{\theta}) = 0$, we obtain that
\begin{equation}
	\bm \eta = \frac{\sum_i \exp_\phi'(\Psi(\bm{\theta})+ \bm{\theta} \cdot  \bm{E}_{i}) \cdot \bm{E}_{i}}{\sum_i \exp_\phi'(\Psi(\bm{\theta})+ \bm{\theta}  \bm{E}_{i})} = \bm{E}_i \cdot \bm{P}^\phi \quad,
\end{equation}
where $\bm{P}^\phi$ is the so-called  \emph{escort distribution}.  With $\exp_\phi'(\log_\phi(x)) = \phi(x)$, the elements of $\bm{P}^\phi$ are given by
\begin{equation}
	P^\phi_j = \frac{\phi(p_j)}{\sum_i \phi(p_i)} =  \frac{\phi(p_j)}{h_\phi(\bm{p})} \quad ,
\end{equation}
where we define $h_\phi(\bm{p}) \equiv\sum_i \phi(p_i)$.
The Legendre transform provides a connection between the exponential family and the escort family, where the coordinates are obtained in the form of escort distributions. This generalizes the results for the ordinary exponential family, where the dual coordinates form a mixture family, which can be obtained as the superposition of the original distribution.
The importance of dual coordinates in information geometry comes from the existence of a dually-flat geometry for the pair of coordinates. This means that there exist two affine connections with vanishing coefficients (Christoffel symbols). For the exponential family, the connection determined by the exponential distribution is called $e$-connection, and the dual connection leading to a mixture family that is called $m$-connection \cite{ay17}. For more details, see e.g. \cite{amari12}. We next look at generalizations of the Kullback-Leibler divergence and the Fisher information for the case of $\phi$-deformations.

\section{Deformed divergences, entropies, and metrics}
For the $\phi$-deformed exponential family we have to define the proper generalizations of the relevant quantities, such as the entropy, divergence, and metric. A natural approach is to start with the deformed Kullback-Leibler divergence, denoted by $D_\phi(\bm{p} \| \bm{q})$. $\phi$-entropy can then be defined as
\begin{equation}
	S_\phi(\bm{p}) \sim - D_\phi(\bm{p} \| \bm{u}_n) \quad ,
\end{equation}
where $\sim$ means that the relation holds up to a multiplicative constant depending only on $n$.
Similarly, $\phi$-deformed Fisher information is
\begin{equation}
	g_{\phi,ij}(\bm{\theta}) = \left. \frac{\partial^2 D_{\bm{\phi}}(p(\bm{\theta}_0) \|  p(\bm{\theta}))}{\partial \theta_i \partial \theta_j} \,  \right |_{\bm{\theta}
	=\bm{\theta}_0}  \quad .
\end{equation}
There is now more than one way to generalize the Kullback-Leibler divergence.  The first is Csisz\'{a}r's divergence
\cite{Csiszar91}
\begin{equation}
	I_f(\bm{p} \| \bm{q}) = \sum_i q_i f(p_i/q_i) \quad ,
\end{equation}
where $f$ is a convex function. For $f(x) = x \ln x$, we obtain the Kullback-Leibler divergence.
Note however, that the related information geometry based on the generalized Fisher information is trivial, because we have
\begin{equation}
	g^f_{ij}(\bm{p}) = f''(1) g_{ij}^F(\bm{p}) \quad ,
\end{equation}
i.e., the rescaled Fisher information metric; see \cite{naudts04}.
The second possibility is to use the divergence of Bregman type, usually defined as
\begin{equation}
D_f(\bm{p} \| \bm{q}) = f(\bm{p}) - f(\bm{q}) - \langle \nabla f(\bm{q}),\bm{p}-\bm{q} \rangle \quad .
\end{equation}
It can be understood as the first-order Taylor expansion of $f$ around $\bm{q}$, evaluated at $\bm{p}$. Let us next discuss two possible types of the Bregman divergence, which naturally correspond to the $\phi$-deformed family. For both, the $\phi$-exponential family is obtained from the maximum entropy principle of the corresponding $\phi$-entropy, however, under different constraints. Note that the maximum entropy principle is just a special version of the more general minimal relative entropy principle, which minimizes the divergence functional $D(\bm{p} \| \bm{q})$ w.r.t. $\bm{p}$, for some given \emph{prior} distribution $\bm{q}$.

\subsection{Linear constraints: divergence a l\`{a} Naudts}
One generalization of Kullback-Leibler divergence was introduced by Naudts \cite{naudts02} by considering $f(\bm{p}) = \sum_i \left(\int_1^{p_i} \log_\phi(x) \ud x + (1-p_i)\right)$, which leads to
\begin{equation}
	D^N_\phi(\bm{p} \| \bm{q}) =\sum_j \int_{q_j}^{p_j}\ud x  \left(\log_\phi(x) - \log_\phi(q_j)\right) \quad .
\end{equation}
The corresponding entropy can be expressed as
\begin{equation}\label{eq:naudtsent}
	S^N_\phi(\bm{p}) = - \sum_j \int_0^{p_j} \ud x \ln_\phi(x) \quad .
\end{equation}
$S^N_\phi(\bm{p})$ is maximized by the $\phi$-exponential family under {\em linear} constraints. The Lagrange functional is
\begin{equation}
	\mathcal{L}_\phi(\bm{p}) = S^N_\phi(\bm{p}) - \Psi \sum_i p_i - \sum_j \theta_j \sum_i p_i E_{ij} \quad ,
\end{equation}
which leads to
\begin{equation}-\log_\phi(p_i) - \Psi - \sum_j \theta_j E_{ij} =0 \quad ,
\end{equation}
and we get
\begin{eqnarray}\label{eq:psi2}
\Psi(\bm{\theta}) &=& -\sum_i p_i \log_\phi(p_i) - \sum_j \theta_j \langle E_j \rangle  \nonumber \\
&=& - \langle \log_\phi(\bm{p})\rangle - \sum_j \theta_j \langle E_j \rangle \quad ,
\end{eqnarray}
which is just Eq. \eqref{eq:psi}, averaged over the distribution $p_i$. Note that Eq. \eqref{eq:psi2} provides the connection to thermodynamics, because $\Psi(\theta)$ is a so-called \emph{Massieu function}. For a canonical ensemble, i.e., one constraint on the average energy, $\theta$, plays the role of an inverse temperature, and $\Psi$ can be related to the free energy, $F(\theta) =  \theta \Psi(\theta)$. Thus, the term $ \langle \log_\phi(\bm{p})\rangle$ can be interpreted as the \emph{thermodynamic entropy}, which is determined from Eq. \eqref{eq:psi2}. This is a consequence of the Legendre structure of thermodynamics. 

The corresponding MaxEnt distribution can be written in the form
\begin{eqnarray}
p_i(\bm{\theta}) &=& \exp_\phi\left(-\langle \log_\phi(\bm{p}) \rangle - \sum_j \theta_j \left(\langle E_j \rangle - E_{ij}\right) \right) \nonumber \\ &=& \exp_\phi\left(\Psi(\bm{\theta})+ \bm{\theta} \cdot \bm{E}_{i}\right) \quad .
\end{eqnarray}
Finally, Fisher information metric can be obtained in the following form
\begin{equation}
g^{N}_{\phi,ij}(\bm{p}) = \log_\phi'(p_i) \delta_{ij}+\log_\phi'(p_0) = \frac{1}{\phi(p_i)}\delta_{ij}+\frac{1}{\phi(p_0)} \quad .
\end{equation}

\subsection{Escort constraints: divergence a l\`{a} Amari}
Amari et al. \cite{amari10,amari12} use a different divergence introduced in \cite{vigelis13}, which is based on the choice, $f(\bm{p}) = \sum_i P_i^\theta \log_\phi(p_i)$.  This choice is motivated by the fact that the corresponding entropy is just the dual function of $\Psi(\bm{\theta})$, i.e., $\varphi(\bm{\eta})$. This is easy to show, because
\begin{eqnarray}\label{eq:varphi}
\varphi(\bm{\eta}) &=& \bm{\eta} \cdot \bm{\theta} -  \Psi(\bm{\theta}) = \sum_j P_j^\phi (\theta_j E_{ij} - \Psi(\bm{\theta})) \nonumber \\ &=& \sum_j  P_j^\phi \log_\phi(p_j) \quad .
\end{eqnarray}
Thus, the divergence becomes
\begin{equation}
D^A_\phi(\bm{p} \| \bm{q}) = \frac{1}{h_\phi(\bm{p})} \sum_j \phi(p_j) (\log_\phi(p_j) - \log_\phi(q_j)) \quad ,
\end{equation}
and the corresponding entropy can be expressed from Eq. \eqref{eq:varphi} as
\begin{equation}
S^A_\phi(\bm{p}) = -\frac{1}{h_\phi(\bm{p})} \sum_j \phi(p_j) \log_\phi(p_j) \quad,
\end{equation}
 so it is a dual function of $\Psi(\bm{\theta})$. For this reason, the entropy is called \emph{canonical}, because it is obtained by the Legendre transform from the Massieu function $\Psi$. Interestingly, the entropy is maximized by the $\phi$-exponential family under {\em escort constraints}. The Lagrange function is
\begin{equation}
	\mathcal{L}_\phi(\bm{p}) = \mathcal{S}^A_\phi(\bm{p}) -  \Psi \sum_i p_i - \sum_j \theta_j \sum_i P_i^\phi E_{ij} \quad .
\end{equation}
After a straightforward calculation we get
\begin{equation}
\Psi(\bm{\theta}) = -\sum_i \phi(p_i(\bm{\theta})) \quad ,
\end{equation}
and the corresponding MaxEnt distribution can be expressed as
\begin{eqnarray}
p_i(\bm{\theta}) &=& \exp_\phi\left(-\langle \log_\phi(\bm{p}) \rangle_\phi - \sum_j \theta_j \left( \langle E_j \rangle_\phi -  E_{ij}\right)\right) \nonumber \\ &=& \exp_\phi\left(\Psi(\bm{\theta})+\bm{\theta} \cdot \bm{E}_{i}\right) \quad ,
\end{eqnarray}
where
\begin{equation}
\Psi(\bm{\theta}) = -\langle \log_\phi(\bm{p}) \rangle_\phi - \sum_j \theta_j \langle E_j \rangle_\phi \quad .
\end{equation}
Here $\langle \cdot \rangle_\phi$ denotes the average under the escort probability measure, $\bm{P}^\phi$. Interestingly, in the escort constraints scenario, the ``MaxEnt'' entropy is the same as the ``thermodynamic'' entropy in the case of linear constraints. We call  this entropy, $S^A_\phi(\bm{p})$, the
\emph{dual entropy.}
Finally, one obtains the corresponding metric
\begin{eqnarray}
g^{A}_{\phi,ij}(\bm{p}) &=&  - \frac{1}{h_\phi(\bm{p})} \left(\frac{\log_\phi''(p_i)}{\log_\phi'(p_i)} \delta_{ij}+ \frac{\log_\phi''(p_0)}{\log_\phi'(p_0)}\right) \nonumber \\ &=&  \frac{1}{h_\phi(\bm{p})}\left( \frac{\phi'(p_j)}{\phi(p_j)} \delta_{ij} + \frac{\phi'(p_0)}{\phi(p_0)}\right) \quad .
\end{eqnarray}
Note that the metric can be obtained from $\Psi(\bm{\theta})$ as $g^{A}_{\phi,ij}(\bm{\theta}) = \frac{\partial^2 \Psi(\bm{\theta})}{\partial \theta_i \partial \theta_j}$, which is the consequence of the Legendre structure of escort coordinates \cite{amari12}.
For a summary for the $\phi$-deformed divergence, entropy and metric, see Table \ref{tab:and}.

\begin{table*}[t]
\centering
\begin{tabular}{l c c c}
  \hline
 & $\phi$-deformation & linear constraints & escort constraints \\
  \hline
divergence  &$D_\phi(\bm{p} \| \bm{q})$ & $\sum_j \int_{q_j}^{p_j}\ud x \left(\log_\phi(x) - \log_\phi(q_j)\right)$ & $\frac{\sum_j \phi(p_j) (\log_\phi(p_j) - \log_\phi(q_j))}{\sum_k \phi(p_k)}$ \\
entropy &   $S_\phi(\bm{p})$ & $-\sum_i \int_{0}^{p_i} \log_\phi(x) \ud x$ & $-\sum_i \phi(p_i) \log_\phi(p_i)/ \sum_k \phi(p_k)$ \\
metric &   $g^\phi_{ij}(\bm{p})$ &  $\frac{1}{\phi(p_i)} \delta_{ij}+\frac{1}{\phi(p_0)}$ &  $\frac{1}{\sum_k \phi(p_k)} \left(\frac{\phi'(p_i)}{\phi(p_i)} \delta_{ij}+\frac{\phi'(p_0)}{\phi(p_0)}\right)$ \\
  \hline
\end{tabular}
\caption{$\phi$-deformation of divergence, entropy and Fisher information corresponding to $\phi$-exponential family under linear and escort constraints. For the ordinary logarithm, $\phi(x)=x$, the two entropies become Shannon entropy, and the divergence is Kullback-Leibler.}
\label{tab:and}
\end{table*}

\subsection{Cram\'{e}r-Rao bound of Naudts type}
One of the important applications of the Fisher metric is the so-called Cram\'{e}r-Rao bound, which is the lower bound for the variance of an unbiased estimator. The generalization of the Cram\'{e}r-Rao bound for two families of distribution was given in \cite{naudts02,naudts04}. Assume two families of distributions, denoted by $\bm{p}(\bm{\theta})$ and $\bm{P}(\bm{\theta})$, with corresponding expectation values, $\langle \cdot \rangle_{\bm{p}(\bm{\theta})}$, and $\langle \cdot \rangle_{\bm{P}(\bm{\theta})}$. Let $c_k$ denote the estimator of the family $\bm{p}(\bm{\theta})$, that fulfills $\langle c_k \rangle_{\bm{p}(\theta)} = \frac{\partial}{\partial \theta_k} f(\bm{\theta})$, for some function $f$, and let us consider a mild regularity condition $\left \langle \frac{1}{\bm{P}(\bm{\theta})} \frac{\partial}{\partial \theta_k}\bm{p}(\bm{\theta}) \right \rangle_{\bm{P}(\bm{\theta})}=0$. Then,
\begin{equation}
	\frac{\langle c_k c_l \rangle_{\bm{P}(\bm{\theta})}
	- \langle c_k \rangle_{\bm{P}(\bm{\theta})}\langle c_l \rangle_{\bm{P}(\bm{\theta})}}
	{\left(\frac{\partial^2 f(\bm{\theta})}{\partial \theta_k \partial \theta_l}\right)^2}
	\geq \frac{1}{I_{kl}(\bm{\theta})} \quad ,
\label{CR}
\end{equation}
where
\begin{equation}
I_{kl}(\bm{\theta}) = \sum_i \frac{1}{P_i(\bm{\theta})} \frac{\partial p_i(\bm{\theta})}{\partial \theta_k}\frac{\partial p_i(\bm{\theta})}{\partial \theta_l} \quad .
\end{equation}
If $\bm{p}(\bm{\theta})=\bm{p}_\phi(\bm{\theta})$ is the $\phi$-exponential family, in Eq. (\ref{CR}) equality  holds for the escort distribution $\bm{P}(\bm{\theta}) = \bm{P}^\phi(\bm{\theta})$, \cite{naudts11}. It is easy to see that for this case, i.e., for the $\phi$-exponential family and the corresponding escort distribution, the following is true
\begin{equation}
	I^N_{\phi;kl}(\bm{p})  = h_\phi(\bm{p}) g^N_{\phi;kl}(\bm{p}) \quad .
\end{equation}
This provides a connection between the Cram\'{e}r-Rao bound and the $\phi$-deformed Fisher metric. In the next section we show that the Cram\'{e}r-Rao bound can be also estimated for the case of the Fisher metric of ``Amari type''.

\section{The information geometric ``Amari-Naudts'' duality}

In the previous section we have seen that there are at least two natural ways to generalize divergence, such that the $\phi$-exponential family maximizes the associated entropy functional, however, under different constraint types. These two ways result in two different geometries on the parameter manifold. The relation between the metric $g^A_{\phi,ij}$ and $g^N_{\phi,ij}$ can be expressed by the operator, $T$
\begin{equation}
	g^A_{ij}(p) = T(g^N_{ij}(p)) \quad ,
\end{equation}
where
\begin{equation}
	T(g(x)) = - N_g \left(\log g(x)\right)'
\end{equation}
with the normalization factor, $N_g = \sum_{i}1/g(p_i)$. Note that the operator acts locally on the elements of the metric. In order to establish the connection to Cram\'{e}r-Rao bound, let us focus on the transformation of $g^A$.

\subsection{Cram\'{e}r-Rao bound of Amari type}

 The metric of the ``Amari case'' can be seen as a conformal transformation \cite{ohara18} of the metric that is obtained in the ``Naudts case'', for a different deformation of the logarithm. Two metric tensors are connected by a conformal transform if they have the same form, except for the global conformal factor, $\Omega(p)$, which depends only on the point $p$. Our aim is to connect the Amari metric with the Cram\'{e}r-Rao bound and obtain another type of bound for the estimates that are based on escort distributions. For this end, let us consider a general metric of Naudts type, corresponding to $\chi$-deformation,  and a metric of Amari type, corresponding to $\xi$-deformation. They are connected through the conformal transform, which acts globally on the whole metric. The relation can be expressed as
\begin{equation}
	g^N_{\chi,ij}(\bm{p}) = \Omega(\bm{p}) g^A_{\xi,ij}(\bm{p}) \quad .
\end{equation}
By using previous results in this relation, we obtain
\begin{equation}
	\frac{1}{\chi(p_i)} = \frac{\Omega(\bm{p})}{h_\xi(\bm{p})} \frac{\xi'(p_i)}{\xi(p_i)} \quad ,
\end{equation}
from which we see that  $\Omega(\bm{p}) = h_\xi(\bm{p})$ and $\log_\chi(x) = \log(\xi(x))$, i.e.,
\begin{equation}
	\xi(x) = \exp(\log_\chi(x)) \Rightarrow \log_\xi(x) = \int_1^x \exp(-\log_\chi(y)) \ud y
\end{equation}
Note that $\log_\chi$ might not be concave because
\begin{equation}
	\frac{\ud^2}{\ud x^2} \log_\chi(x) = \frac{\xi(x) \xi''(x) - \xi'(x)^2}{\xi(x)^2} \quad .
\end{equation}
Concavity exists, if $\xi''(x) \leq \frac{\xi'(x)^2}{\xi(x)}$. To now make the connection with the Cram\'{e}r-Rao bound, let us take $\chi(x) = \phi(x)$, so $\xi(x) = \exp\log_\phi(x)$, and
\begin{equation}
	I^A_{\phi;kl}(\bm{p})  = h_{\exp(\log_\phi)}(\bm{p}) g^A_{\exp(\log_\phi);kl}(\bm{p}) \quad .
\end{equation}
As a consequence, there exist two types of Cram\'{e}r-Rao bounds for a given escort distribution, which might be used to estimate the lower bound of the variance of an unbiased estimator, obtained from two types of Fisher information.

\subsection{Example: Duality of $(c,d)$-entropy}

We demonstrate the ``Amari-Naudts''  duality on the general class of $(c,d)$-entropies \cite{ht11a,ht11b}, which include all deformations associated to statistical systems that fulfil the first three Shannon-Khinchin axioms. These include most of the popular deformations, including  Tsallis $q$-exponentials \cite{tsallis88}, and stretched exponentials studied in connection with entropies by Anteneodo and Plastino \cite{anteneodo99}. The generalized $(c,d)$-logarithm is defined as
\begin{equation}
	\log_{(c,d)}(x) = r- r x^{c-1} \left(1-\frac{(1-(1-c) r)}{d r} \log x\right)^d \quad ,
\end{equation}
where $c$ and $d$ are the scaling exponents \cite{ht11a,thk-book}, and $r$ is a free scale parameter (that does not influence the asymptotic behavior). The associated $\phi$-deformation is
\begin{eqnarray}
& &	\phi_{(c,d)}(x) = \nonumber \\ && \frac{x}{r-\log_{c,d}(x)} \left(\frac{ (-c r+r-1) \log (x)+d r}{(c-1) ((c-1) r+1) \log (x)+d}\right)
 \, .
\end{eqnarray}
The inverse function of $\log_{(c,d)}$, the deformed $(c,d)$-exponential, can be expressed in terms of the Lambert W-function
\begin{eqnarray}
	&& \exp_{(c,d)}(x) =  \nonumber \\ &&\exp\left(-\frac{d}{1-c} \left[W\left(B (1-x/r)^{1/d}\right) - W(B)\right]\right) \, ,
\end{eqnarray}
where $B = \frac{(1-c)r}{1-(1-c)r}\exp\left(\frac{(1-c)r}{1-(1-c)r}\right)$.
The corresponding entropy that is maximized by $(c,d)$-exponentials (see \cite{thk-book} for their properties), is $(c,d)$-entropy
\begin{equation}
	S_{(c,d)}(\bm{p}) = r A^{-d} e^A \sum_i \Gamma(1+d,A-c\ln p_i)- r c \quad ,
\end{equation}
where $A = \frac{cdr}{1-(1-c)r}$. This is an entropy of ``Naudts type'', since it is maximized with $(c,d)$-exponentials under {\em linear} constraints. We can immediately write the metric as
\begin{eqnarray}
	&& g^N_{(c,d),ij}(\bm{p}) = \nonumber \\ && \left(\frac{r-\log_{c,d}(p_i)}{p_i}\left( \frac{(c-1) ((c-1) r+1) \log (p_i)+d}{ (-c r+r-1) \log (p_i)+d r}\right)\right) \delta_{ij}\nonumber\\ &+& \frac{r-\log_{c,d}(p_0)}{p_0} \left( \frac{(c-1) ((c-1) r+1) \log (p_i)+d}{ (-c r+r-1) \log (p_0)+d r}\right)
\, .
\end{eqnarray}

The corresponding entropy of ``Amari type'', i.e., maximized with $(c,d)$-exponentials under the {\em escort} constraints
\begin{equation}
	\frac{\sum_i \phi_{(c,d)}(p_i) E_i}{\sum_j \phi_{(c,d)}(p_j)} = \langle E \rangle_{(c,d)} \quad ,
\end{equation}
is
\begin{widetext}

\begin{equation}
	S_{(c,d)}^A(\bm{p}) = - \frac{1}{h_{(c,d)}(\bm{P})} \sum_i \frac{p_i \log_{c,d} p_i}{r-\log_{c,d}(p_i)} \left(\frac{ (-c r+r-1) \log (p_i)+d r}{(c-1) ((c-1) r+1) \log (p_i)+d}\right)
 \quad ,
\end{equation}
\end{widetext}

and its metric finally is
  \begin{widetext}

	\begin{eqnarray}
	g_{(c,d),ij}^A(\bm{p}) =  \frac{1}{p_i} \left(2-c-\frac{(d-1) ((c-1) r+1)}{((c-1) r+1) \log (p_i)-d r}-\frac{(c-1)^2
   r+c-1}{(c-1) d r-c d r+(c-1) ((c-1) r+1) \log (p_i)+d+d r}\right) \, \delta_{ij} \nonumber\\
   + \frac{1}{p_0} \left(2-c-\frac{(d-1) ((c-1) r+1)}{((c-1) r+1) \log (p_0)-d r}-\frac{(c-1)^2
   r+c-1}{(c-1) d r-c d r+(c-1) ((c-1) r+1) \log (p_0)+d+d r}\right) \quad .
\end{eqnarray}
\end{widetext}


\begin{figure}[t]
\includegraphics[width=5cm]{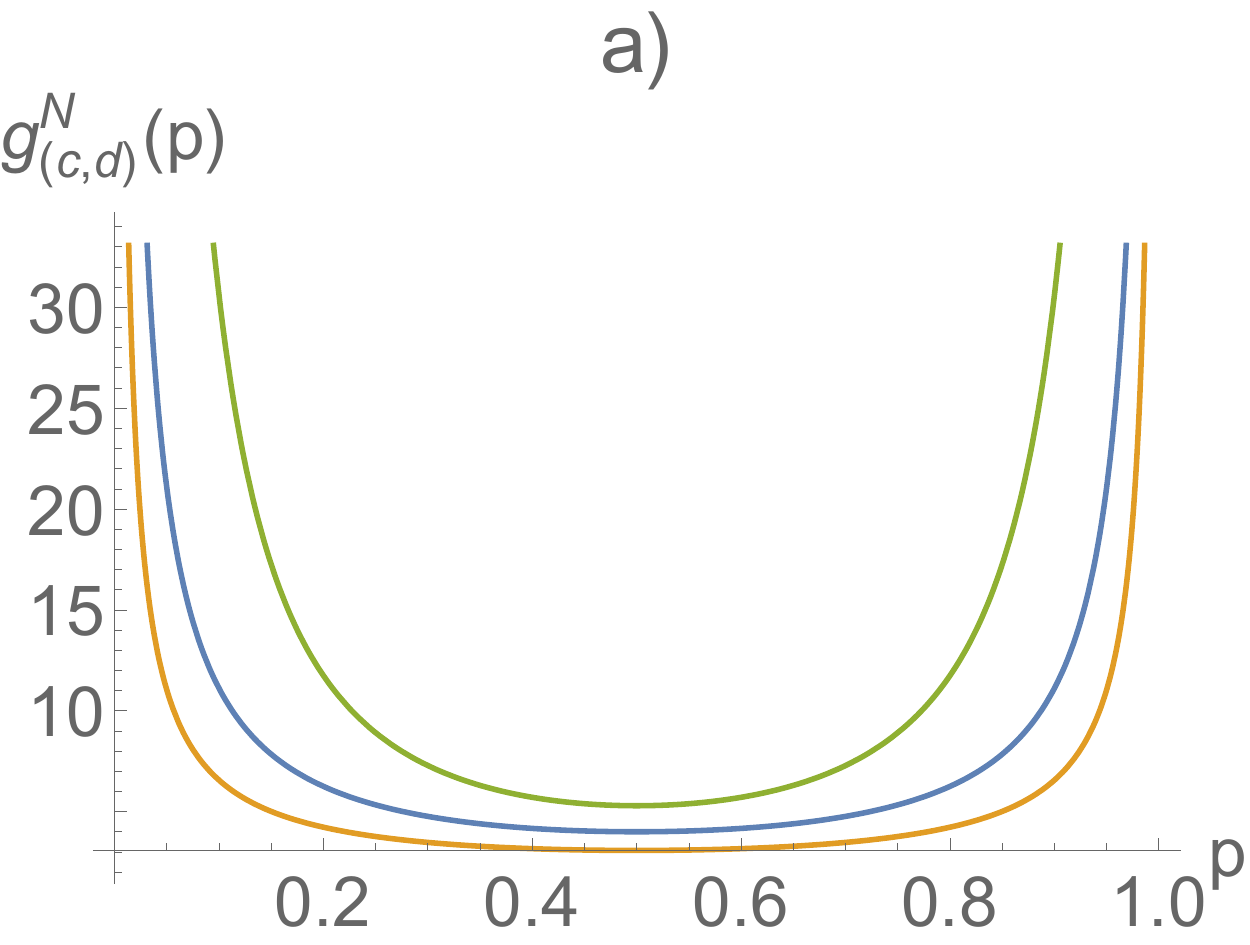}
\includegraphics[width=5cm]{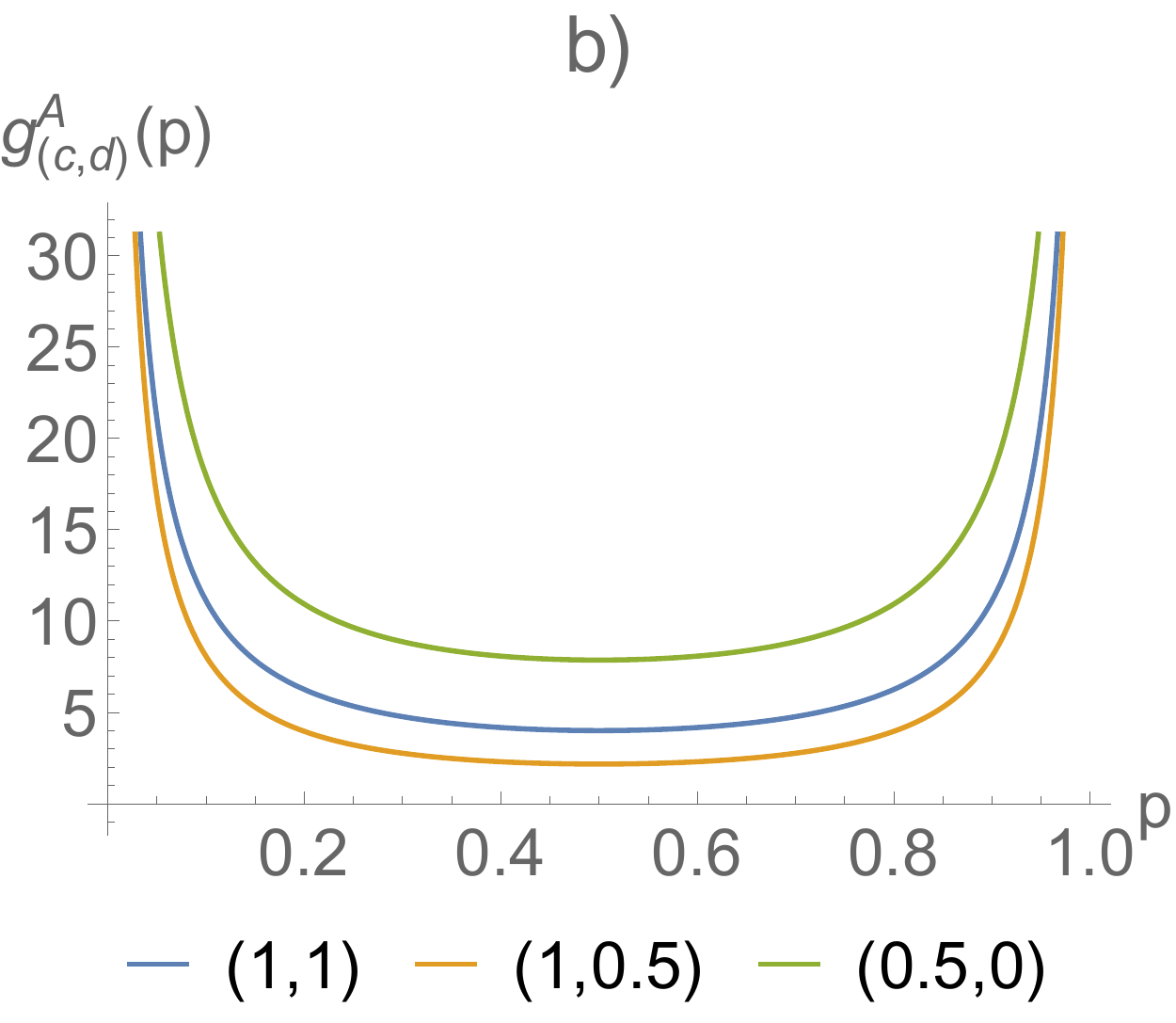}
\includegraphics[width=5cm]{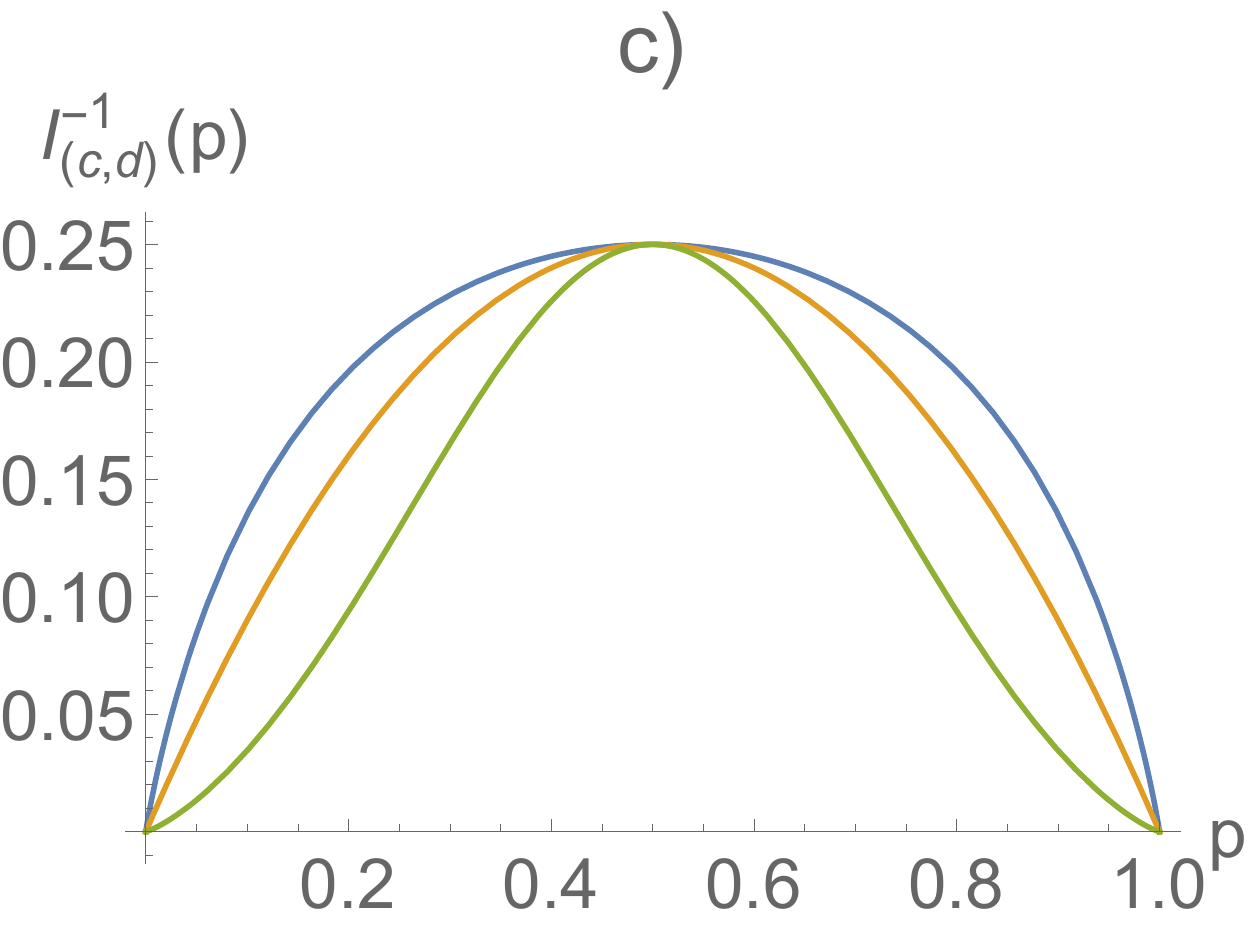}
\caption{Fisher metric for $p=(p,1-p)$ corresponding to  various $(c,d)$-deformations ($(c,d)=(1,1)$, $(1,1/2)$, $(1/2,0)$) for (a) Naudts type,
(b) Amari type, and (c) the Cram\'{e}r-Rao bound corresponding to the metric.}
\label{fig:g}
\end{figure}

\begin{figure}[t]
\includegraphics[width=8cm]{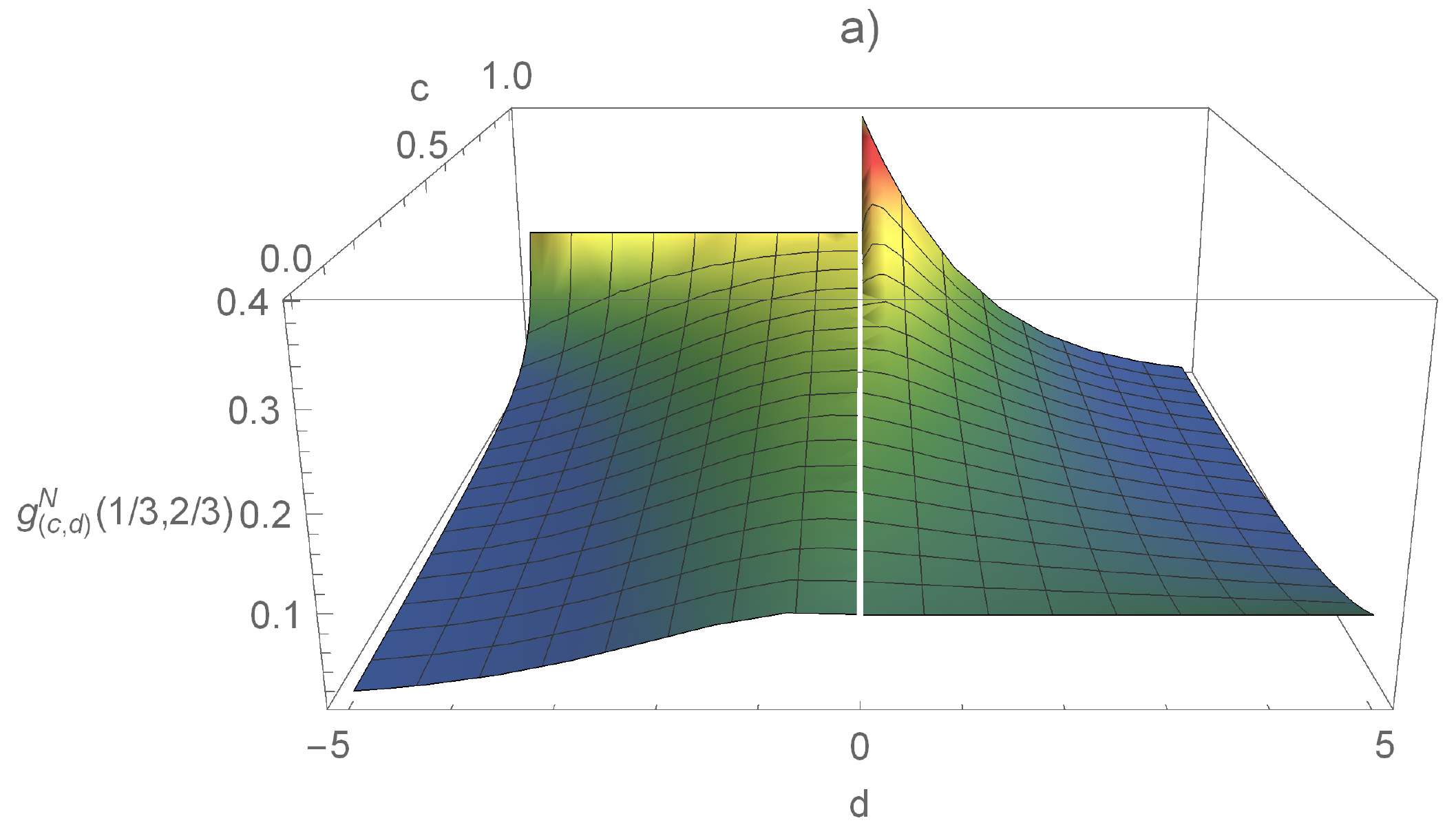}
\includegraphics[width=8cm]{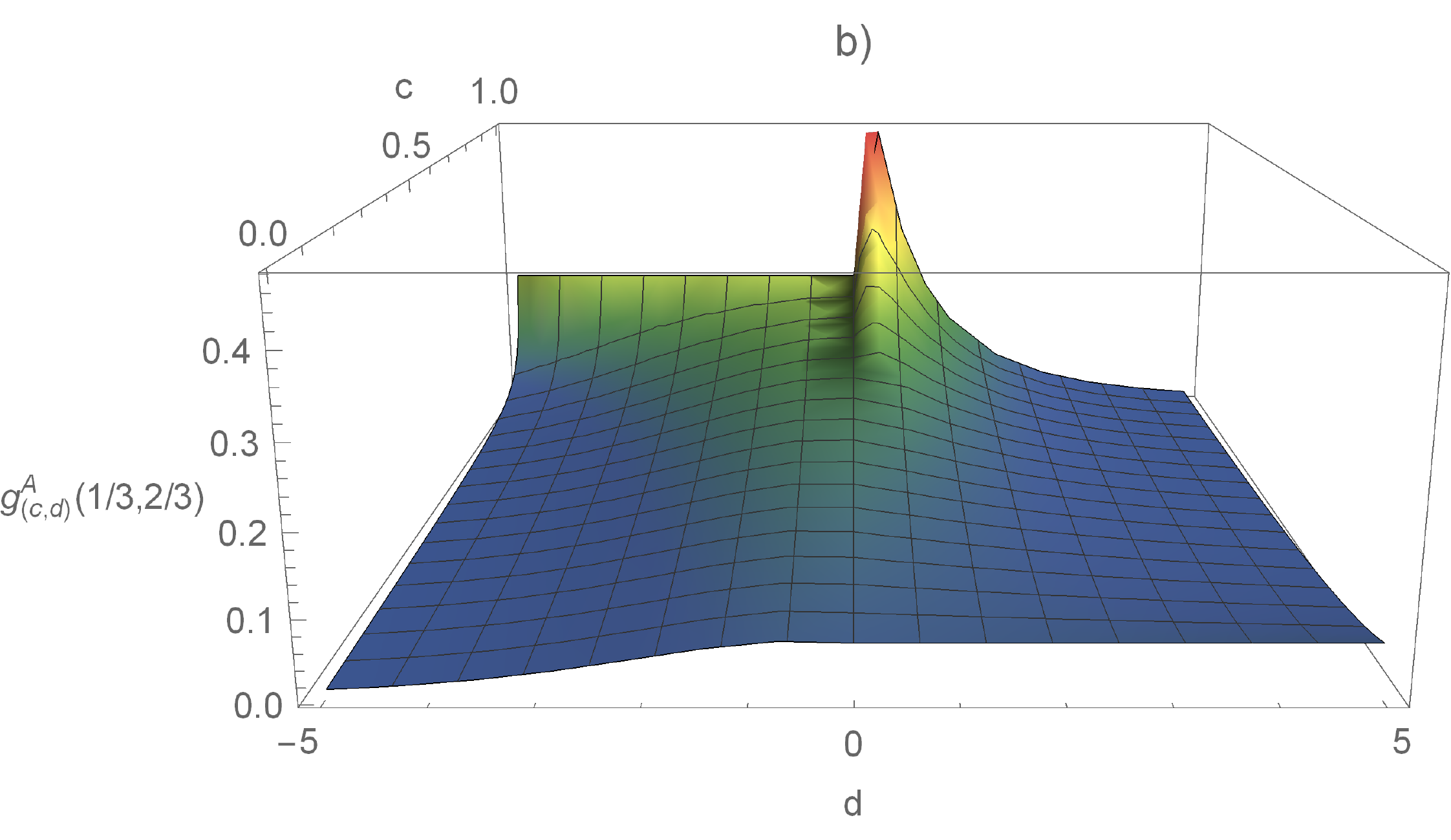}
\caption{Fisher metric for $(c,d)$-deformations as a function of $c$ and $d$ of Naudts type (a), and Amari type (b). The metric is evaluated at a point $p = (1/3,2/3)$.  \label{fig:s}}
\end{figure}

The metric of Amari type for the $(c,d)$-entropy was already discussed in \cite{ghikas17} based on $(c,d)$-logarithms. However, as demonstrated above, the metric can be found without using the inverse $\phi$-deformed logarithms, which in the case of $(c,d)$-logarithms lead to Lambert $W$-functions. The Fisher metric of Naudts and Amari type and the corresponding Cram\'{e}r Rao bound is shown in Fig. \ref{fig:g}. The scaling parameter is set (following \cite{ht11a}) to, $r = 1/(1-c+cd)$, for $d \leq 0$, and $r = \exp(-d)/(1-c)$, for $d < 0$. The Fisher metric of both types is displayed in Fig. \ref{fig:s}  as a function of the parameters $c$ and $d$ for a given point, $P=(1/3,2/3)$. We see that both types of metric have a singularity for $(c,d) = (1,0)$. This point corresponds to distributions with compact support. For one-dimensional distributions the singularity corresponds to the transition between distributions with support on the real line and distributions with support on a finite interval.

Interestingly, for $(c,d)=(q,0)$, the metric simplifies to
\begin{equation}
	g_{(q,0),ij}^A(\bm{p}) = \frac{2-q}{p_i} \delta_{ij} +\frac{2-q}{p_0} \quad ,
\end{equation}
which corresponds to Tsallis $q$-exponential family. Therefore, $g_{(q,0),ij}^A(p)$ is just a conformal transformation of the Fisher information metric for the exponential family, as shown in \cite{amari12}. Note, that only for Tsallis $q$-exponentials the relation between $S_q^N(p)$ and $S_q^A(p)$ can be expressed as (see also Table \ref{tab:phi})
\begin{equation}
	S_q^A(\bm{p})= f(S_{q'}^N(\bm{p})) \quad ,
\end{equation}
where $f(x)=  (2-q)/x$ and $q'=2-q$. This is nothing but the well-known additive duality $q \leftrightarrow 2-q$ of Tsallis entropy \cite{tsallis05}. Interestingly, $q$-escort distributions form a group with $\phi_q(\phi_{q'}(x)) = (\phi_{q \cdot q'}(x))$ and $\phi^{-1}_q(x) =\phi_{1/q}(x)$, where $q \leftrightarrow 1/q$ is the multiplicative duality \cite{tsallis17}. This is not the case for more general deformations, because typically, the inverse does not belong to the class of escort distributions. Popular deformations belonging to the $(c,d)$-family, as Tsallis $q$-exponential family or the stretched exponential family are summarized in Table \ref{tab:phi}.

\section{Connection to the deformed-log duality}

A different duality of entropies and their associated logarithms under linear and escort averages has been discussed in \cite{htg12}. There, two approaches were discussed. The first is an approach using generalized entropy of trace form under linear constrains.  It was denoted by 
\begin{equation}
	S^{HT}(\bm{p}) = \sum_i s^{HT}(p_i) = \sum_i \int_0^{p_i} \log_{}^{HT}(x) \ud x \quad. 
\label{HT}
\end{equation}
It corresponds to the Naudts case here, $\log_{}^{HT}(x)=\log_{\phi}^{N}(x)$. The second approach, originally introduced by Tsallis and Souza  \cite{tsallis03}, uses the trace form entropy
\begin{equation}
	S^{TS}(\bm{p}) = \sum_i s^{TS}(p_i) = \sum_i \int_0^{p_i} \log^{TS}(x) \ud x \quad ,
\label{TS}
\end{equation}
which is maximized under the escort constraints
\begin{equation}
	\langle E \rangle_{TS} = \frac{\sum_j u(p_j) E_j}{\sum_j u(p_j)} \quad ,
\end{equation}
where $u(p_j) = p_j + \nu s_{TS}(p_j)$. The linear case is recovered for $\nu=0$. This form is dictated by the Shannon-Khinchin axioms, as discussed in the next section. Let us assume that the maximization of both approaches, Eq. (\ref{HT}) under linear, and Eq. (\ref{TS}) under escort constraints leads to the same MaxEnt distribution. One can then show that there exists the following duality (deformed-log duality) between $\log^{HT}(x)$ and $\log^{TS}$(x)
\begin{equation}
	\log^{TS}(x) = \frac{1}{\frac{1}{\log^{HT}(x)}+\nu}
\end{equation}
Let us focus on specific $\phi$-deformations, so that $\log^{HT}(x) = \log^{HT}_{\phi}(x)\equiv \log^{N}_{\phi}(x)$. Then, $\log^{TS}(x)$ is also a $\phi$-deformation, with
\begin{eqnarray}
	\phi^{TS}(x) &=& \phi^{HT}(x) \left(1 + \nu \log_\phi^{HT}(x)\right)^2 \nonumber \\ &=& \phi^{HT}(x) \left(1 + \nu \int_1^x \frac{1}{\phi^{HT}(y)}\ud y\right)^2
\end{eqnarray}
It is straightforward to calculate the metric corresponding to the entropy $S^{TS}(\bm{p})$
\begin{equation}
	g_{\phi,ij}^{TS}(\bm{p}) = \frac{1}{\phi^{TS}(p_i)} \delta_{ij} + \frac{1}{\phi^{TS}(p_0)}\nonumber\\
	= \mathcal{T}_\nu(g_{\phi,ij}^{HT}(\bm{p})) \quad ,
\end{equation}
where
\begin{equation}
\mathcal{T}_\nu(g(x)) = g(x)\left(1 + \nu \int_1^x g(y) \ud y \right)^2 \quad .
\end{equation}
Thus, the Tsallis-Souza approach results yet in another information matrix. We may also start from the other direction and look at the situation, when the escort distribution for the information geometric approach is the same as the escort distribution for the Tsallis-Souza approach. In this case we get that
\begin{equation}
	\phi(x) = x + \nu s^{TS}(x) = x + \int_0^x \mathcal{T}_{\nu}(\log_\phi(y)) \ud y \quad .
\end{equation}
We find that the entropy must be expressed as
\begin{equation}
	S_\phi(\bm{p}) = \sum_i \frac{\phi(p_i) - p_i}{\nu}
\end{equation}
Note that for $\phi(x) = x^q$, and $\nu=1-q$, we obtain Tsallis entropy
\begin{equation}
	S_\phi(\bm{p}) = \sum_i \frac{p_i^q - p_i}{1-q} \quad ,
\end{equation}
which corresponds to $S^{TS}(\bm{p})$ for $q'=2-q$, which is nothing but the mentioned Tsallis additive duality. It turns out  that Tsallis entropy is the only case where the deformed-log duality and the information geometric duality result in the same class of functionals. In general, the two dualities have different escort distributions.

\begin{table}[t]
\centering
\begin{tabular}{l l l  }
  \hline
 & Tsallis $q$-exponential &  Stretched $\eta$-exponential  \\
    \hline
$\phi(x)$			& $x^q$ 					 &  $ x \eta \log(x)^{1-1/\eta}$\\
$\log_\phi(x)$		& $\frac{x^{1-q}-1}{1-q}$ 		 & $\log(x)^{1/\eta}$ \\
$\exp_\phi(x)$ 		& $\left(1+ (1-q) x\right)^{1/(1-q)}$  & $ \exp(x^\eta)$ \\
$\chi_\phi(x)$		& $\frac{x}{q}$	&  $\frac{x \eta \log(x)}{(\eta-1) +\eta \log(x)}$ \\
$S_\phi^N(\bm{p})$ & $\frac{1}{q-1} \left(\sum_i \frac{p_i^{2-q}}{2-q} - 1\right)$ & $\sum_i \Gamma\left(\frac{\eta+1}{\eta},-\log p_i\right)$ \\
$S_\phi^A(\bm{p})$ & $\frac{1}{1-q}\left( \frac{1}{\sum_{i} p_i^q}-1\right)$ & $\frac{\sum_{i} p_i \log p_i}{\sum_i p_i (\log p_i)^{1-1/\eta}}$ \\
  \hline
\end{tabular}
\caption{Two important special cases of $(c,d)$-deformations and related quantities: Power laws (Tsallis \cite{tsallis88}) and stretched exponentials \cite{anteneodo99}.}
\label{tab:phi}
\end{table}

\section{Discussion}
In this paper discuss the information geometric duality of entropies that are maximized by $\phi$-exponential distributions under two types of constraint: linear constraints that are known from contexts such as thermodynamics, and escort constraints, that appear naturally in the theory of statistical estimation and information geometry. This duality implies two different entropy functionals:
$S^N(\bm{p}) = -\sum_i \int_0^{p_i} \ud x \log_\phi(x)$, and
$S^A(\bm{p}) = -\langle \log_\phi(P)\rangle_\phi$. For $\phi(x)=x$, they both boil down to Shannon entropy. The connection between the entropy of Naudts type and the one of Amari type can be established through the corresponding Fisher information through the Cram\'{e}r-Rao bound. Contrary to the deformed-log duality introduced in \cite{htg12}, the information theoretic duality introduced here cannot be established within the framework of $\phi$-deformations, since $S^A(\bm{p})$ is not a trace form entropy. We demonstrated the duality between the Naudts approach with linear constraints, and the Amari approach with escort constraints, with the example of $(c,d)$-entropies, which include a wide class of popular deformations, including Tsallis and Anteneodo-Plastino entropy as special cases. Finally, we compared in detail the information geometric duality to the deformed-log duality, and showed that they are fundamentally different, and result in other types of Fisher information.

Let us now discuss the role of information geometric duality and possible applications in information theory and thermodynamics.  Recall that the Shannon entropic functional is determined by the four Shannon-Khinchin (SK) axioms. In many different contexts at least three of the axioms should hold
\begin{itemize}
\item (SK1) Entropy is a continuous function of the probabilities $p_i$ only, and should not explicitly depend on any other parameters.
\item (SK2) Entropy is maximal for the equi-distribution ${p}_{i}=1/W$.
\item (SK3) Adding a state $W+1$ to a system with ${p}_{W+1}=0$ does not change the entropy of the system.
\end{itemize}
The fourth axiom that describes the composition rule for entropy (originally for Shannon entropy, $S(A + B) = S(A) + S(B|A)$). The only entropy satisfying all four SK axioms is Shannon entropy. However, Shannon entropy is not sufficient to describe statistics of complex systems \cite{ht11b}, and can lead to paradoxes in applications in thermodynamics \cite{jensen18}. Therefore, instead of imposing the fourth axiom in situations where it does not apply, it is convenient to consider a weaker requirement, such as generic scaling relations of  entropy in the thermodynamic limit \cite{ht11a,korbel18}. It is possible to show that the only type of duality satisfying the first three Shannon-Khinchin axioms is the deformed-log duality of \cite{htg12}. Moreover, entropies which are neither trace-class, nor sum-class (i.e., in the form $f(\sum_i g(p_i)$) might be problematic from the view of information theory and coding. For example, it is then not possible to consistently introduce a conditional entropy \cite{ilic13} because the corresponding conditional entropy cannot be properly defined. This is related to the fact that the Kolmogorov definition of conditional probability is not generally valid for escort distributions \cite{jizba17}. Additonal issues arise from the theory of statistical estimation, since only sum-class entropies can fulfil the consistency axioms \cite{uffink95}. From this point of view, the deformed-log duality using the class of Tsallis-Souza escort distributions can play the role in thermodynamical applications \cite{htg11}, because the corresponding entropy fulfils the SK axioms. 
On the other hand, the importance of escort distributions considered by Amari and others is in realm of information geometry (e.g., dually flat geometry or generalized Cram\'{e}r-Rao bound), and their applications in thermodynamics might be limited. Finally, for the case of Tsallis $q$-deformation both dualities, the information geometric and the deformed-log duality reduce to the well-known additive duality $q \leftrightarrow 2-q$.


This work was supported by the Austrian Science Fund FWF under I 3073.
All authors contributed to the conceptualization of the work, the discussion of the results, and their interpretation.  
JK took the lead in technical computations. JK and ST wrote the manuscript.



\bibliographystyle{plain}

\end{document}